\documentclass[referee]{raa}            

\usepackage{graphicx,times}             
\usepackage{natbib}
\usepackage{amssymb,amsmath}
\bibpunct{(}{)}{;}{a}{}{,}

\usepackage[a4paper=true,dvipdfm=true,pagebackref=true]{hyperref}
\hypersetup{colorlinks = true, linkcolor = green, anchorcolor = red, citecolor = blue, filecolor = red, pagecolor = red, urlcolor = red}
\begin{document}

\title{Distance to the Dorado galaxy group
}
\setcounter{page}{1}          
\author{N.A.~Tikhonov\inst{1}, O.A.~Galazutdinova\inst{1}}

	   \institute{Special Astrophysical Observatory, Nizhnij Arkhyz, Karachai-Cherkessian Republic, 
		Russia 369167; {\it ntik@sao.ru}\\
}

\abstract
{Based on the archival images of the Hubble Space Telescope, stellar photometry of the
 brightest galaxies of the Dorado group: NGC\,1433, NGC\,1533, NGC\,1566 and NGC\,1672 was carried out. 
 Red giants were found on the obtained CM diagrams and distances to the galaxies were measured
 using the TRGB method. The obtained values: $14.2\pm 1.2$, $15.1\pm 0.9$, $14.9 \pm 1.0$ and $15.9\pm 0.9$~Mpc,
  show that all the named galaxies are located approximately at the same distances and form a scattered 
  group with an average distance $D = 15.0$~Mpc. It was found that blue and red supergiants are visible
   in the hydrogen arm between the galaxies NGC\,1533 and IC\,2038, and form a ring structure in the lenticular
    galaxy NGC\,1533, at a distance of 3.6~kpc from the center. The high metallicity of these stars ($Z = 0.02$)
     indicates their origin from NGC\,1533 gas.
}

\authorrunning{N.A.~Tikhonov, O.A.~Galazutdinova}            
\titlerunning{Distance to the Dorado galaxy group}  

\maketitle

\section{Introduction}           
A concentration of galaxies of different types and luminosities can be observed in the 
southern constellation Dorado. Among them, \citet{Shobbrook_1966} identified 11 galaxies, 
which, in his opinion, constituted one group, which he called ``Dorado''. Based on the 
 measured radial velocities and photometry of the galaxies, \citet{Shobbrook_1966} estimated 
 the distance to the group as 9.8~Mpc. Six massive galaxies: NGC\,1433, NGC\,1533, NGC\,1549,
  NGC\,1553, NGC\,1566, and NGC\,1672, around which fainter galaxies are concentrated, as clearly
 seen in the diagram of \citet{Kilborn_etal_2005}, determine the position of the group in the 
 celestial sphere. The scattered position of the Dorado galaxies has led to various hypotheses
  for their unification into groups. \citet{Vaucouleurs_1975} combined the bright Dorado galaxies into
 groups G16, G21 and G22 and estimated the distance to the main group G16 as 18.4~Mpc. In the
  same year, \citet{Sandage_1975} published lists of galaxy groups, where the Dorado group consisted of
   12 main and 6 probable members. It was indicated that the distance to the group is ~16.9~Mpc. Subsequently, the Dorado group changed its composition several times, depending on the criteria for selecting galaxies or after obtaining new data on radial velocities or distance measurements. \citet{Huchra_Geller_1982} divided 28 Dorado galaxies into two groups, HG3 and HG8, and \citet{Maia_etal_1989} increased the number of Dorado galaxies to 60, placing them in groups 7 and 13. \citet{Huchra_Geller_1982} and \citet{Maia_etal_1989} estimated distances to the galaxy groups based on the radial velocities and the Hubble constant $H = 100$. According to their estimates, all Dorado galaxies were located no further than 12~Mpc.
   
From the photographic observations of the group, \citet{Ferguson_Sandage_1990} carried out photometry of the galaxies up to $B = 20\,\!^{\rm m}$ and included 79 galaxies in the Dorado group. For most of the new galaxies, the radial velocities were unknown; therefore, the enrollment in the group was based on the location of faint galaxies in relation to the bright ones. Based on the deep images, \citet{Carrasco_etal_2001} found 69 low surface brightness galaxies in the Dorado group and determined their color indices $(V-I)$ and magnitudes in the filters $V$ and $I$.

Studies have shown that the Dorado group is one of the richest groups of galaxies in southern sky. Four bright galaxies, stellar photometry of which we will present below, have active nuclei, despite their different morphological types. NGC\,1433 and NGC\,1672 belong to Sy2 galaxies, NGC\,1566 to Sy1, and NGC\,1533 to LINER galaxies (NED). Therefore, the issue of measuring the distances to the Dorado galaxies requires a solution.

After taking images of the Dorado galaxies with the Hubble Space Telescope, it became possible to determine the distance using the surface brightness fluctuation method (SBF method), which \citet{Tonry_etal_2001} used to measure the distances to several Dorado galaxies. The resulting average value $D = 18.5$~Mpc has put this group further than it was previously thought.

The Tully--Fisher (TF) method is typically used to determine the distances to spiral galaxies. Using this method, \citet{Tully_etal_2009} determined that the distances to NGC\,1433 and NGC\,1672 were 8.32~Mpc and 11.9~Mpc, respectively. In the extensive work of \citet{Tully_etal_2013}, which presents the results of measuring the distances to many galaxies by different methods, the distances to two other Dorado galaxies are indicated as 20.5~Mpc (to NGC\,1533) and 6.61~Mpc (to NGC\,1566). The distance to NGC\,1566 is questionable, since NGC\,1533 has a heliocentric velocity $v_h = 790~{\rm{km\,s^{-1}}}$, while NGC\,1566 has a much higher velocity, $v_h = 1504~{\rm{km\,s^{-1}}}$. However, \citet{Sorce_etal_2014} estimated that the distance to NGC\,1566 is even less: $D = 5.5$--$6.0$~Mpc.

The estimation of distances using the radial velocities demonstrated the values that were less varying from measurement to measurement. \citet{Firth_etal_2006} measured the radial velocities of the galaxies of Dorado group and determined the distance to it as 16.9~Mpc. According to their results, the group itself is not virialized due to the existence of subgroups within the group.

Recent measurements of distances to the galaxies NGC\,1433 and NGC\,1566 by the TRGB method \citep{Sabbi_etal_2018} gave values of 9.04~Mpc and 17.9~Mpc. The distances obtained by \citet{Sabbi_etal_2018} will be discussed in further detail in Section \ref{Res_dis}.

The measurements of distances mentioned above indicate the great uncertainty of derived values, which leads to the same uncertainties in the Dorado group members list. The difficulties are increasing by the fact that the galaxies are significantly scattered across the sky and none of them, thus far, has an exact distance value. It is impossible even to indicate which of the brightest galaxies is closer to us, since distances for them vary widely (Table~\ref{tab1}). For weaker galaxies, the scatter of measurements can be even larger. More realistic distance values can be obtained by simply dividing the radial velocity by the Hubble constant, but this will not take into account the proper velocities of galaxies, the values of which are usually in the range from 50 to 150~${\rm{km\,s^{-1}}}$.  However, within groups and clusters these velocities can be significantly higher because of the interaction of galaxies with each other.
 
\begin{table*}[!]
		\begin{center}
	\caption{Parameters of bright galaxies of the Dorado group} \label{tab1}
	\begin{tabular}{l|c|c|c|c|c|c|c|c} \hline
Galaxy &  T&  $B_t$& $v_h$, ${\rm{km\,s^{-1}}}$ &a$\times$b, arcmin & $D$, Mpc & $M_B$, mag&$D_{\rm{min}}$, Mpc&$D_{\rm{max}}$, Mpc 
		\\
		\hline
		N\,1433 &  (R)SB(r)ab&  10.70 & 1076 & 6.5$\times$5.9 & 14.2 & $-20.08$ & 8.3\,[1]&11.6\,[2]
		\\
		N\,1533 &  SB(rs)0   &  11.70 & 790  & 2.8$\times$2.3 & 15.1 & $-19.26$ & 13.4\,[2]&24.1\,[3]
		\\
		N\,1566 &  SAB(s)bc  &  10.33 & 1504 & 8.3$\times$6.6 & 14.9 & $-20.56$ & 5.5\,[4]&21.3\,[5]
		\\
		N\,1672 &  SB(s)b    &  10.28 & 1331 & 6.6$\times$ 5.5 & 15.9& $-20.81$ & 9.9\,[6]&14.5\,[2]
		\\
		\hline
	\end{tabular}
	\end{center}
\end{table*}

The basic properties of the studied galaxies are listed in Table~\ref{tab1}. The results of the classification of galaxies (T), the values of heliocentric velocities ($v_h$), the apparent magnitudes ($B_t$) and the sizes of galaxies in arc minutes (a$\times$b) given in Table ~\ref{tab1} are taken from the NED, while the measurements of distances and luminosities of the galaxies was performed by us. The minimum and maximum distances are taken from the studies: [1]~--- \citet{Tully_etal_2009}, [2]~--- \citet{Tully_Fisher_1988}, [3]~--- \citet{Springob_etal_2014}, [4]~--- \citet{Sorce_etal_2014}, [5]~--- \citet{Willick_etal_1997}, [6]~--- \citet{Giraud_1985}.

To avoid the difficulties in compiling a list of galaxies of one large group, it can be divided into subgroups, which include those galaxies that are concentrated around the bright ones. This is how the groups  NGC\,1533, NGC\,1566, NGC\,1672 \citep{Kilborn_etal_2005} appeared, which previously belonged to the same Dorado group. Something similar is observed in the Virgo cluster of galaxies, where groups of galaxies around M\,87, M\,86 and M\,49 are distant from each other, but together they make up one Virgo cluster.

\section{Stellar photometry}

The images that could be used for stellar photometry and distance determination for four bright galaxies of the Dorado group were obtained on the Hubble Space Telescope under different programs and in different years. Unfortunately, most of the images were taken in the central regions of galaxies, where the bright background of galaxies and the presence of supergiants, and AGB stars brighter than red giants made these regions unsuitable for determining the distance using the TRGB method. Therefore, we carried out distance measurements only for the stars at the periphery of such galaxies. This selection led to a decrease in the total number of stars in the sample, but increased the quantity of red giants in relation to other types of stars, which made it possible to measure the position of the upper edge of the red giant branch (TRGB jump), needed to calculate the distance.

\begin{table*}[!]
	\begin{center}
		\caption{The Hubble Space Telescope data archive} \label{tab2}
		\begin{tabular}{l|c|c|c|c|c|c|c|c} \hline
			Galaxy & ID & Camera & F814W & F606W  & F555W & F435W & F110W  & F160W
			\\
			\hline
			NGC1433   & 13364 & WFC3 & 986  & -- & 1140& -- & --&--
			\\
			NGC1433   & 12659 & WFC3 & --   & -- & -- & -- & 1212 & 1412
			\\
			NGC1533-1 & 10438 & ACS & 4950  &2288& -- & -- & -- & --
			\\
			NGC1533-2 & 10438 & ACS & 4950  &2288& -- & -- & -- & --
			\\
			NGC1566   & 13364 & WFC3 & 989  & -- & 1143& -- & -- & --
			\\
			NGC1566   & 12999 & ACS & 1244  & -- & 768& -- & -- & --
			\\
			NGC1672   & 10354 & ACS & 2444  & -- & -- & 2444 & -- & --
			\\
			NGC1672   & 15654 & ACS & 3775  & 3063 & -- & -- & -- & --
			\\
			\hline
		\end{tabular}
	\end{center}
\end{table*}

To study the stellar composition of galaxies and determine the distance, we used the archival images of the Hubble Space Telescope, obtained on proposal ID 10438, 10354, 12659, 12999, 13364 and 15654 with the ACS/WFC and WFC3 cameras. Table~\ref{tab2} contains information on the original photometric data of the Hubble Space Telescope: proposal ID, camera (ACS or WFC3), filter F814W ($I$), F606W ($V$), F555W ($V$), F435W ($B$), F110W ($IR$) F160W ($IR$) with exposure in seconds.
\begin{figure*}
	\centerline{\includegraphics[width=0.8\linewidth]{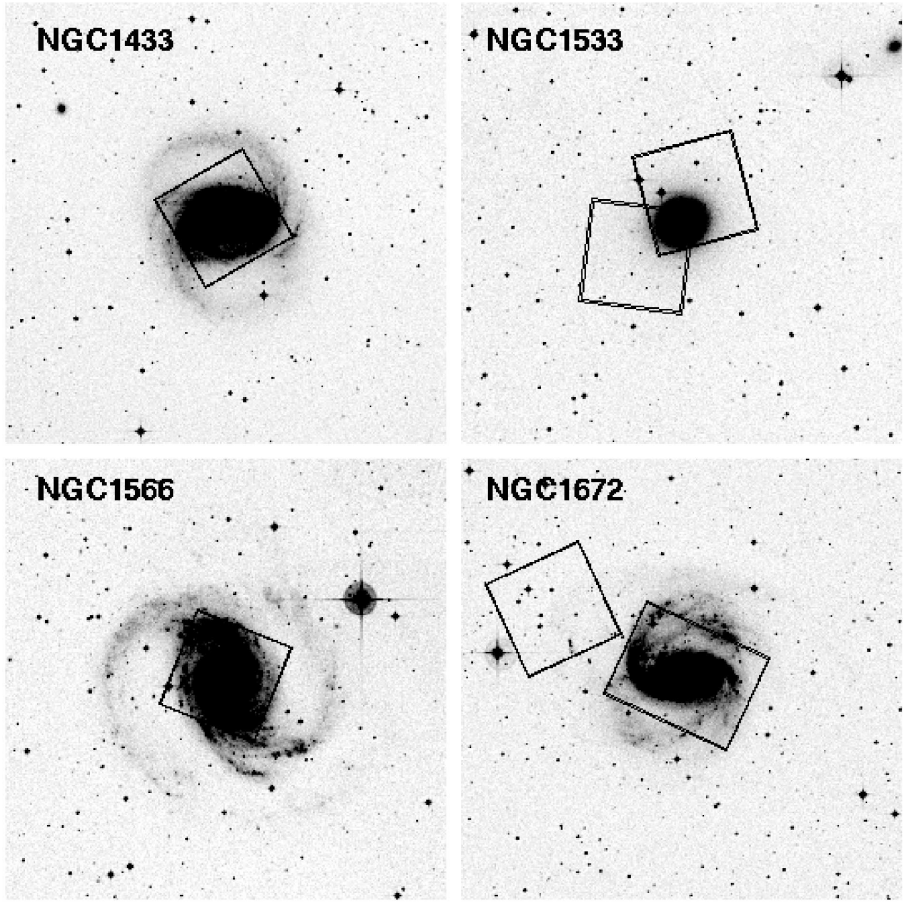}}
	\caption{Images of the galaxies of Dorado group in the DSS survey images in the $B$ filter.
		 The rectangles mark the fields of the Hubble Space Telescope. The size of the image of each galaxy is $15\arcmin \times 15\arcmin$, north is at the top.}
	\label{fig1}
\end{figure*}
\begin{figure*}
	\centerline{\includegraphics[angle=0, width=14cm,clip]{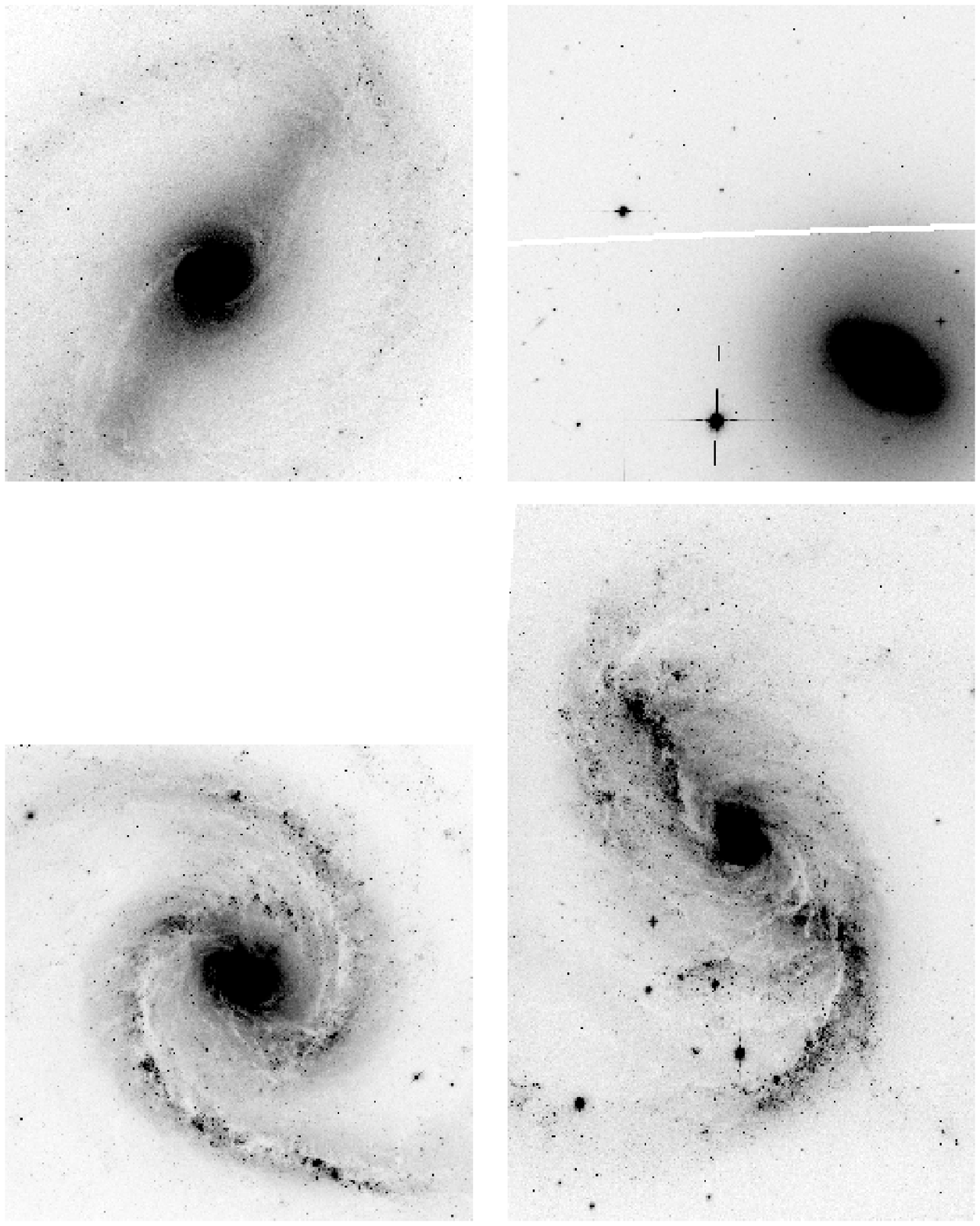}}
	\caption{HST images of galaxies shown in Fig.~\ref{fig1}: filter F555W (NGC\,1433 and NGC\,1566), F606W (NGC\,1533) and F435W (NGC\,1672).}
	\label{fig2}
\end{figure*}
 
Figure~\ref{fig1} shows the DSS (Digitized Sky Survey) images of four main galaxies of the Dorado group in blue filter $B$ with marked of the HST fields, and Fig.~\ref{fig2} shows images of the same galaxies obtained with the HST telescope on the F606W, F555W and F435W filters.

Stellar photometry of the galaxies was carried out using two software packages: DAOPHOT~II \citep{Stetson_1987,Stetson_1994} and DOLPHOT~2.0\footnote {http://americano.dolphinsim.com/dolphot/dolphot.pdf}. The photometry of stars with both programs was carried out in a standard way. For DAOPHOT~II, this was described in our previous study \citep{Tikhonov_etal_2009,Tikhonov_Galazutdinova_2009}. The DOLPHOT~2.0 package was used in accordance with the recommendations of \citep{Dolphin_2016}, and the photometry procedure consisted of preliminary masking of bad pixels, removal of traces of cosmic particles, and further PSF photometry of the stars found in two filters. To remove diffuse objects: star clusters, distant or compact galaxies, all stars were selected according to the $C\!H\!I$ and $S\!H\!A\!R\!P$ parameters, which determine the shape of the photometric profile of each measured star \citep{Stetson_1987}. The profiles of diffuse objects differed from the profiles of isolated stars that we selected as standard, which made it possible to carry out such a selection with the lists of stars obtained by DAOPHOT~II and DOLPHOT~2.0.

\begin{figure*}
	\centerline{\includegraphics[width=0.8\linewidth]{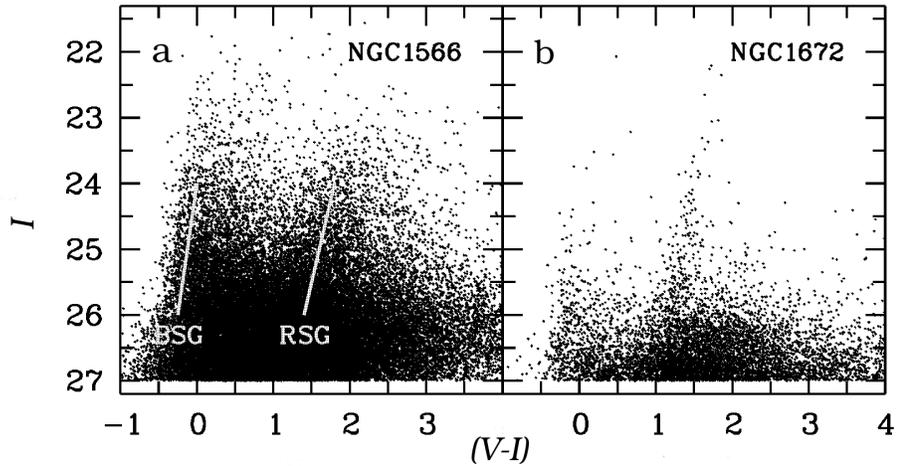}}
	\caption{CM diagram of the stars of central regions of the galaxy NGC\,1566 and the periphery of the galaxy NGC\,1672. The lines mark the position of the blue supergiant (BSG) and red supergiant (RSG) branches in NGC\,1566.}
	\label{fig3}
\end{figure*}
The principles of photometry with DOLPHOT and DAOPHOT are the same, but there are some differences in their use. For example, in DAOPHOT we took single stars from the studied fields as PSF stars, and in DOLPHOT we used the PSF profile library. The difference in the results obtained with these programs is noticeable when comparing the apparent distribution of very faint stars over the image field. Due to the ineffectiveness of charge transfer and the existence of residual traces of cosmic particles, DOLPHOT shows an excessive number of faint stars in the central region of the field, while the distribution of stars obtained with DAOPHOT is closer to real. However, when there is a large concentration of stars, there is a problem of choosing PSF stars in DAOPHOT. With the pros and cons of both software packages in mind, we used both of them and compared the results. When measuring the positions of TRGB jumps, both methods gave similar results and no significant differences were found between them.

The Hertzsprung--Russell diagrams (CM diagrams) of three spiral galaxies obtained by photometry of stars are common diagrams for this type of galaxies, thus, as an example, we present the NGC\,1566 diagram and the NGC\,1672 periphery diagram (Fig.~\ref{fig3}). The diagrams clearly show the branches of blue and red supergiants. Red giants do not stand out visually due to the large number of brighter supergiants and AGB stars.
\begin{figure*}
	\centerline{\includegraphics[width=0.82\linewidth]{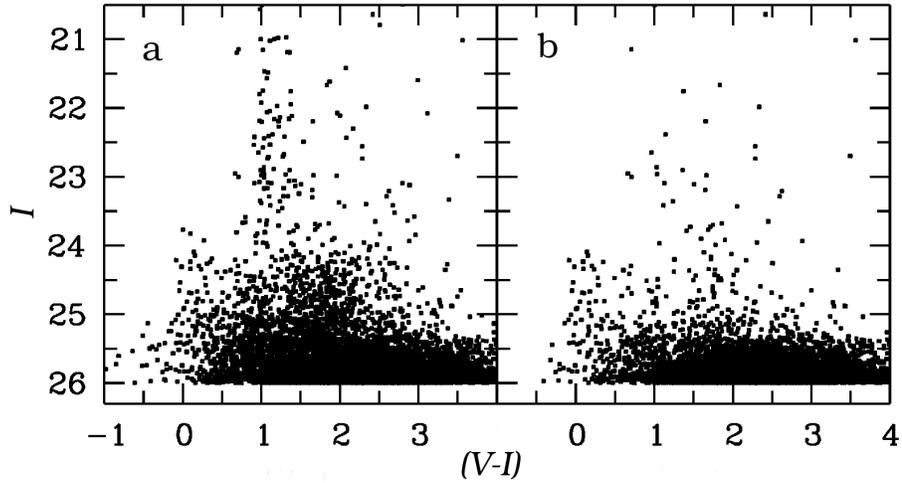}}
	\caption{CM diagram of the galaxy NGC\,1533 at different $C\!H\!I$ values. At $C\!H\!I <2.5$, globular clusters fall into the diagram, the branch of which is visible at $(V-I) = 1$. At $C\!H\!I <1.2$, stars and very few compact clusters remain on the diagram.}
	\label{fig4}
\end{figure*}

The diagram of the galaxy NGC\,1533 (Fig.~\ref{fig4}) is interesting, because in the lenticular galaxy, where large regions of star formation are not visible, there are blue and red supergiants. Radio observations at H\,I \citep{Ryan-Weber_etal_2004} revealed a ring structure around NGC\,1533 and an extended arm connecting it to the IC\,2038 dwarf galaxy, with which it interacts \citep{Cattapan_etal_2019}. The structure of the gas arm between galaxies is clearly visible in the study of \citet{Werk_etal_2010} (diagram 7).
The obtained CM diagram of NGC\,1533 (Fig.~\ref{fig4}a) shows a branch inhabited by globular clusters at a color index $(V-I) = 1$. A similar diagram is presented in the work of \citet{DeGraaff_etal_2007}. However, the diagram in Fig.~\ref{fig4}a was obtained with an deliberately increased value of the parameter $C\!H\!I <2.5$, which led to the appearance of diffuse objects in the list of stars and in the CM diagram. If we use the standard value $C\!H\!I <1.2$, then most globular clusters will disappear from the CM diagram, and only stars and very compact star-shaped globular clusters will remain (Fig.~\ref{fig4}b).

\begin{figure*}[!]
	\centering{\includegraphics[width=0.8\linewidth]{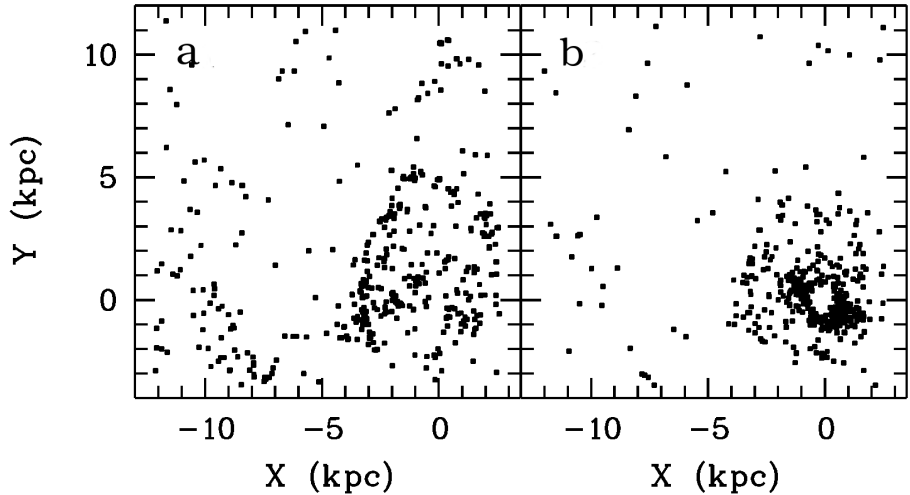}}
	\caption{Distribution of blue (a) and red (b) supergiants in the form of a ring in the galaxy NGC\,1533. The center of the coordinate system is aligned with the center of the galaxy. The size and orientation of the diagram correspond to the size of the NGC\,1533 image in Fig.~\ref{fig2}. The concentration of stars at the galactic periphery belongs to the gas arm between the galaxies NGC\,1533 and IC\,2038.}
	\label{fig5}
\end{figure*}
The blue stars, the branch of which is visible in the diagram in Fig.~\ref{fig4}, are partially scattered throughout the body of the galaxy, but most of them are part of small clusters located mainly in the inner regions of the galaxy. Single clusters are visible even beyond 30~kpc from the galactic center \citep{DeGraaff_etal_2007, Werk_etal_2008, Werk_etal_2010}. An important question in the study of young stars in NGC\,1533 ~ is the source of gas from which the young stars, visible in the CM diagram, were born. Since NGC\,1533 interacts with the dwarf irregular galaxy IC\,2038, which has hydrogen, it can be assumed that the periphery of the dwarf galaxy could have been stripped off and became a source of hydrogen for NGC\,1533.

The diagram in Fig.~\ref{fig5}a shows the distribution of blue stars with a color index $(V-I) <0.3$ over the body of the galaxy. It can be seen that individual stars and small clusters form a ring around the center of the galaxy. In addition, there is a concentration of these blue stars in the periphery, where the HST image captures a fragment of the gas arm between NGC\,1533 and IC\,2038. The CM diagram of NGC\,1533 (Fig.~\ref{fig4}b) shows a low-contrast branch of red stars with a color index $(V-I) = 1.75$. We have selected these stars and plotted their distribution over the body of the galaxy (Fig.~\ref{fig5}b). It can be seen that these red stars, like blue supergiants, also form a ring and are concentrated in the arm between the two galaxies. The presence of a ring structure from regions of star formation is confirmed by images from the space telescope GALEX in the near and far ultraviolet, which show separate clusters, also forming a ring structure, like young stars. It can be noted that in the morphological description of the type of galaxy NGC\,1533~--- SB(rs)0, the presence of a ring is also indicated, however, it is unclear whether the ring structure of young stars that is visible in our diagrams is related to the ring in morphological description.

\begin{figure}
	\centerline{\includegraphics[width=0.5\linewidth]{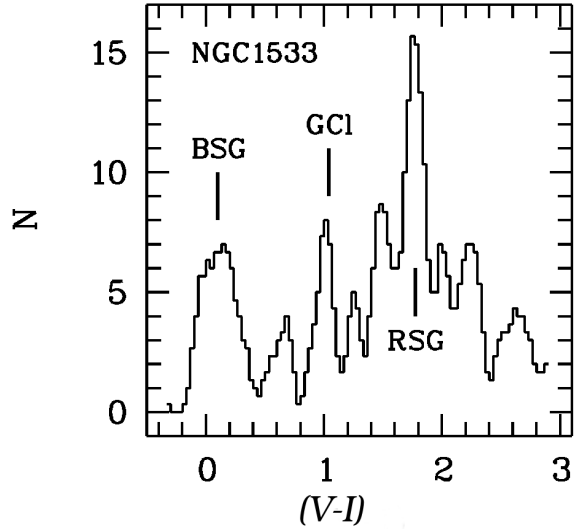}}
	\caption{Distribution of stars by color index in the galaxy NGC1533 in the region of inner ring structure at $2.6<Rad<4.4$~kpc. The diagram shows peaks corresponding to blue supergiants (BSG), red supergiants (RSG) and compact globular clusters (GCl).}
	\label{fig6}
\end{figure}
To study the stellar composition of the ring visible in Fig.~\ref{fig5}, we identified the stars included in it at $700<Rad<1200$ pixels, which corresponds to $2.6<Rad<4.4$~kpc. Figure~\ref{fig6} demonstrates the distribution of these stars by color index $(V-I)$. The diagram easily identifies the maxima that correspond to blue supergiants (BSG) and compact globular clusters at $(V-I) = 1$. It is quite obvious that the maximum at $(V-I) = 1.75$ corresponds to red supergiants. It becomes clear why blue and red stars form almost identical apparent distributions over the body of the galaxy. The red supergiants are not very old, and they are concentrated in the same regions where they were born, and where the younger blue supergiants are visible.
\begin{figure}
	\centerline{\includegraphics[width=0.5\linewidth]{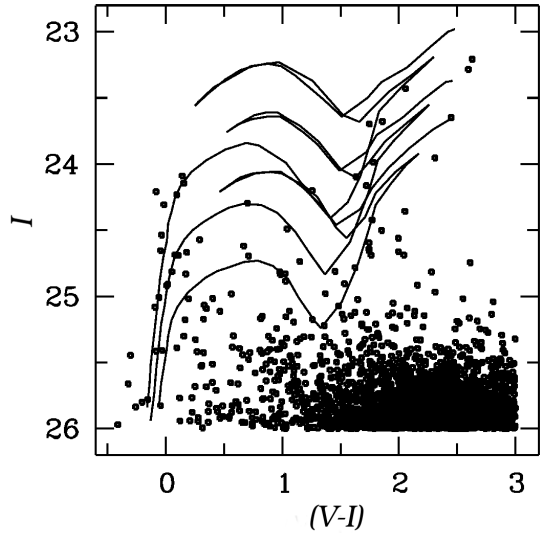}}
	\caption{CM diagram of stars of the galaxy NGC\,1533 in the region of the ring structure at $2.6<Rad<4.4$~kpc. The most suitable isochrones with metallicity $Z = 0.02$ and ages of 12, 22, and 28 Myr are inscribed.}
	\label{fig7}
\end{figure}

Figure~\ref{fig7} shows the CM diagram of the stars that make up the ring in Figure~\ref{fig5}. In this diagram, we have entered the most suitable isochrones from \citet{Bertelli_etal_1994}. These isochrones showed that the age of red supergiants is in the range from 12 to 30 million years, and their metallicity is equal to that of the Sun ($Z = 0.02$). It is possible that older supergiants are also present in the sample, but they are difficult to identify. The obtained results indicate that the hydrogen, from which the stars of the ring were formed, cannot be the hydrogen of the dwarf galaxy IC\,2038, whose luminosity is $M_V  = -16$, since such galaxies have a lower metallicity of hydrogen clouds and young stars born from them \citep{Tikhonov_2018}. In addition, the high metallicity indicates that the young stars visible in NGC\,1533 cannot be the stars of the first wave of star formation at the galactic periphery, as suggested by \citet{Ryan-Weber_etal_2004}.

\begin{figure*}[!]
	\centerline{\includegraphics[width=0.8\linewidth]{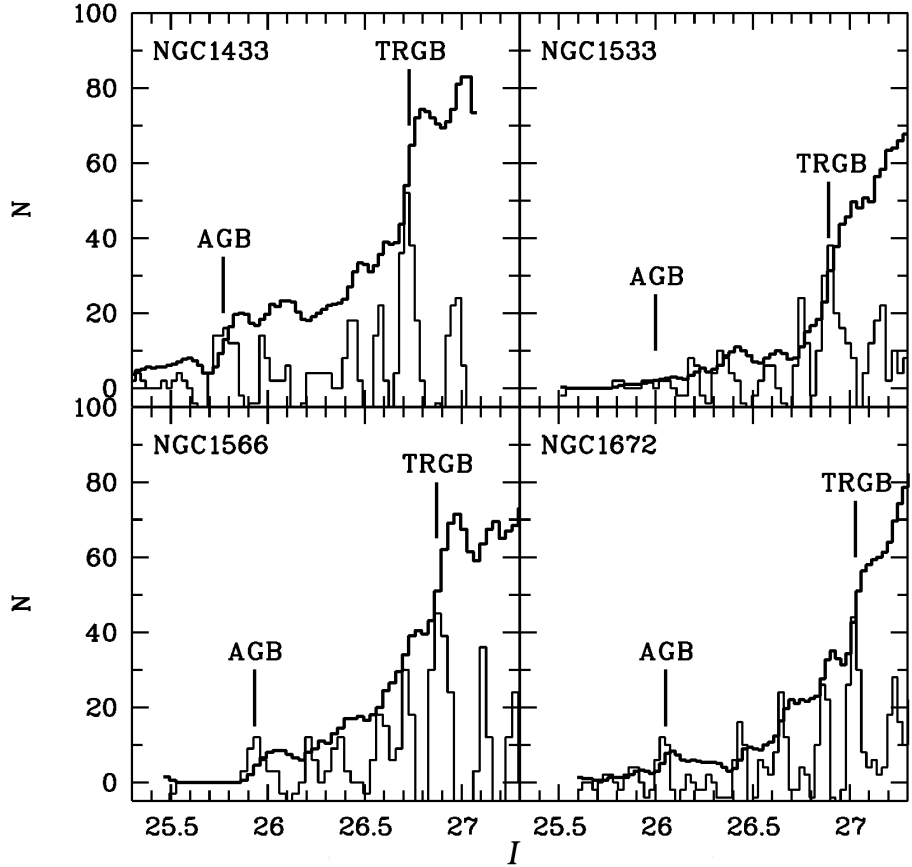}}
	\caption{ Luminosity function of red giants and AGB stars for four galaxies. The vertical lines mark the positions of TRGB jumps and the boundaries of the increase in the number of AGB stars.}
	\label{fig8}
\end{figure*}
Red supergiants are also visible in the arm between the galaxies NGC\,1533 and IC\,2038, although in smaller numbers than in the ring. Their metallicity value is slightly lower than that for the stars of the ring. It can be assumed that the decrease in metallicity of stars in the arm between the galaxies indicates a possible mixing of the gases of two galaxies, one of which supplies hydrogen, and the other enriches it with metals.

Figure~\ref{fig5} shows that blue and red stars are located not only in the ring and arm, but also closer to the center of the galaxy. Indeed, the CM diagram of the central region of galaxy contains single blue and red supergiants at $0.7<Rad<1.7$~kpc, but the main concentration of red stars near the center of the galaxy (Fig.~\ref{fig5}b) is created by AGB stars.

\section{Measuring the distances}

The branch of red giants, which is necessary for measuring the distance, is not visible in the CM diagrams of galaxies (Fig.~\ref{fig3}) due to the large number of brighter stars and the increased brightness of the galaxy body. In the images of the galaxy NGC\,1533 (Fig.~\ref{fig1}), it is easy to find areas in the periphery where bright stars are absent and the galaxy's brightness is low. In such areas, a selection of stars was carried out to determine the TRGB jump. In NGC\,1672, the stellar halo extends far beyond the galaxy's body visible in Fig.~\ref{fig1}, so an image of the galaxy's periphery, where bright stars occupy only a part of the image, was used to search for red giants. It was more difficult to create selections of stars near the galaxies NGC\,1433 and NGC\,1566, where almost the entire area is occupied by bright stars with an increased background brightness. For these galaxies, the stars were chosen outside the spiral arms at $Rad>2000$~px, which corresponds to $100^{\prime\prime}$ or 6.9~kpc for NGC\,1433 and $Rad>2500$~px, which is equal to $125^{\prime\prime}$ or 9.0~kpc for NGC\,1566. In addition, color index selection ($1.2 <(V-I) <1.7$) was used so that the luminosity function of red giants was not affected by blue stars and AGB stars with a high color index.

After the selection, the luminosity functions were obtained for the stars of four galaxies (Fig.~\ref{fig8}). The diagrams show the beginnings of the branches of red giants (TRGB jumps) and AGB stars. The difference between them is approximately one magnitude. For an objective representation of the positions of TRGB jumps, we used the Sobel function \citet{Madore_Freedman_1995}, the maxima of which correspond to sharp changes in the number of stars, which is observed at the border of the red giant branch. In the diagrams in Fig.~\ref{fig8}, the thin line shows the Sobel function, the positions of the maxima of which we used to determine the distances to galaxies.

In addition to TRGB jumps, we measured the  color indices of the red giants $(V-I)_{\rm{TRGB}}$, the values of which do not differ from those of galaxies of this type. The extinction of light towards the galaxies is taken from \citet{Schlafly_Finkbeiner_2011} and is shown in Table~\ref{tab1}.

Using the work of \citet{Lee_etal_1993} on the application of TRGB method, we determined the distances to the galaxies NGC\,1433, NGC\,1533, NGC\,1566, and NGC\,1672. The results are shown in Table~\ref{tab3}. Distance measurement accuracy (external accuracy) specified in Table~\ref{tab3} is the result of the addition of several possible sources of measurement errors. The accuracy of the method of \citet{Lee_etal_1993} is $0\,.\!\!^{\rm m}1$. The accuracy of determining the TRGB jump varies from galaxy to galaxy, but does not exceeds $0\,.\!\!^{\rm m}05$.  The remaining components of the measurement error do not exceed $0\,.\!\!^{\rm m}02$--$0\,.\!\!^{\rm m}03$.

\begin{table*}
		\begin{center}
	\caption{The results of photometry of galaxies} \label{tab3}
	\begin{tabular}{l|c|c|c|c|c|c} \hline
		Galaxy &  $I_{\rm{TRGB}}$, mag&  $(m-M)$, mag &  $D$,~Mpc  & [Fe/H] &  $A_I$, mag  &  $E(V-I)$, mag
		\\
		\hline
		NGC\,1433   &  26.72 &  30.75  & 14.15$\pm$1.15 &$-1.46$ &  0.014 &  0.011
		\\
		NGC\,1533   &  26.89 &  30.90  & 15.12$\pm$0.90 &$-1.64$ &  0.024 &  0.020
		\\
		NGC\,1566   &  26.87 &  30.86  & 14.88$\pm$1.00 &$-1.79$ &  0.014 &  0.011
		\\
		NGC\,1672   &  27.03 &  31.00  & 15.86$\pm$0.92 & $-1.74$ & 0.035 & 0.029
		\\
		\hline
	\end{tabular}
	\end{center}
\end{table*}
	
	\section{Results and discussion}
	\label{Res_dis}
	
	For the first time, the accurate distances for the main galaxies of the Dorado group were obtained using the TRGB method, which show that the subgroups of galaxies around NGC\,1433, NGC\,1566 and NGC\,1672 have approximately the same distances and we can assume that they form a single group in which the virialization process has not ended. The average distance to the Dorado group, based on measurement of four galaxies, without making corrections for the masses of individual galaxies, is $D = 14.99$~Mpc. This is significantly less than the distance measured based on the SBF method and is more consistent with the distance obtained from the radial velocities and the Hubble constant. The measured distances to the galaxies will make it possible to accurately estimate the energy of their active nuclei and to establish the spatial positions of the Dorado galaxies among their neighbors.
	
	It was found that young stars of the lenticular galaxy NGC\,1533  form an inner ring, and are also visible in arm between galaxies  NGC\,1533 and IC\,2038 (Fig.~\ref{fig5}).  The metallicity of these stars equal to that of the Sun ($Z = 0.02$), and their age reaches 30 million years. The high metallicity of the stars of the ring indicates that the hydrogen, from which they were born, can only belong to NGC\,1533, despite the fact that this galaxy is poor in hydrogen. The lower metallicity of the stars in the arm between NGC\,1533 and IC\,2038 probably indicates the participation of the dwarf galaxy IC\,2038, which has low metallicity hydrogen in sufficient quantity, in the processes of star formation in the arm between these galaxies.
	
	In the publications of  \citet{Carrasco_etal_2001} and \citet{Firth_etal_2006} it is reported that the first mention of the Dorado group under the number 18 was from \citet{SH_1957}. However, in this publication, Shahbazyan studies only of the star cluster in the northern sky. Apparently, the first error was copied by later authors without reading the article. However, the \citet{SH_1957} paper is not available on the Internet.
	
	The NED datebase states that \citet{Sabbi_etal_2018} measured the distances to the galaxies NGC\,1433 and NGC\,1566 using the TRGB method. For the galaxy NGC\,1566, this message is erroneous, since the authors themselves indicate that the TRGB jump is outside their CM diagram. For the galaxy NGC\,1433, \citet{Sabbi_etal_2018} obtained the value $D = 9.1$~Mpc, i.e. it should be a relatively close galaxy. Indeed, if we take the stars of this galaxy outside the bright core and carry out their usual  selection, then a jump in the luminosity function can be seen at $I = 25\,.\!\!^{\rm m}7$, which corresponds to the distance $D = 9.1$~Mpc. However, this jump refers to AGB stars, and the actual TRGB jump is weaker by one magnitude (Fig.~\ref{fig8}).  A similar error for the galaxies Maffei1 and Maffei2, where the beginning of the branch of AGB stars was taken as a TRGB jump, was described in detail in our study \citep{Tikhonov_Galazutdinova_2018}.
		
	\begin{figure}
		\centering{\includegraphics[width=0.5\linewidth]{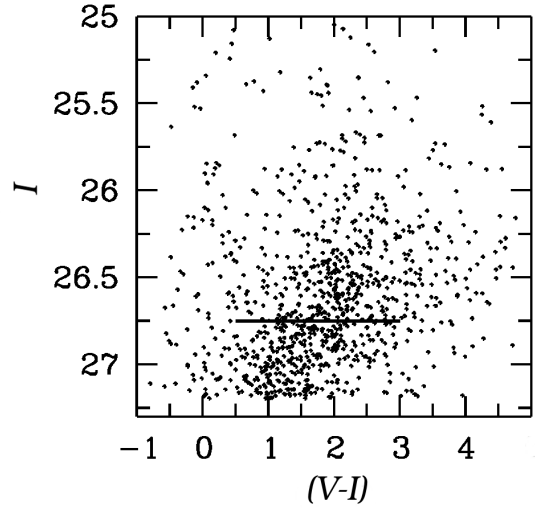}}
		\caption{CM diagram of stars in the periphery of NGC\,1433. The horizontal line marks the position of the TRGB-jump. The apparent clustering of stars above this line is formed by AGB stars in the galaxy disk. At $I = 25\fm7$, which corresponds to $D = 9.1$~Mpc, there is no red giant branch.}
		\label{fig9}
	\end{figure}
	\begin{figure}
		\centering{\includegraphics[width=0.5\linewidth]{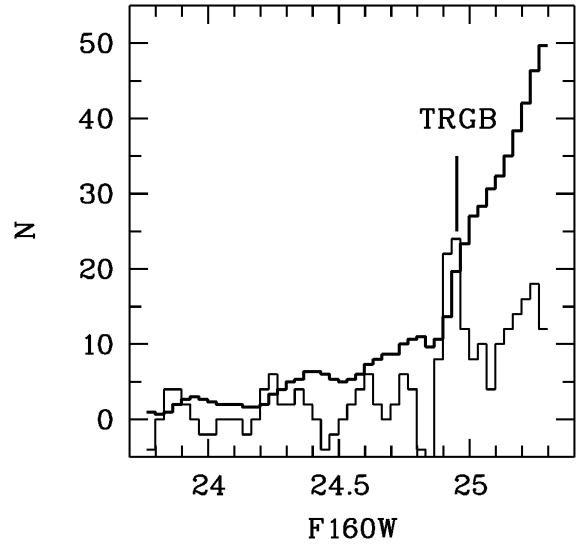}}
		\caption{Function of the luminosity of stars in the infrared region with the F160W filter at a distance of $30\arcmin$ from NGC\,1433. The vertical line shows the position of the TRGB-jump, at F160W = $24\fm95$. This value is consistent with the distance to NGC\,1433 at $I_{\rm{TRGB}} = 26\fm72$.}
		\label{fig10}
	\end{figure}
	
	To show the actual position of the red giant branch, Fig.~\ref{fig9} demonstrates the CM diagram of the periphery of the galaxy NGC\,1433. The diagram shows that at $I = 25\,.\!\!^{\rm m}7$, which corresponds to 9.1~Mpc, the branch of red giants is absent, i.e. the measurement for this galaxy by \citet{Sabbi_etal_2018} is wrong. Visually, it seems that in the diagram in Fig.~\ref{fig9} the TRGB jump is visible at $I = 26\,.\!\!^{\rm m}5$, however, a more detailed study of the distribution of stars shows that the actual TRGB jump is at $I = 26\,.\!\!^{\rm m}72$ (Fig.~\ref{fig8}), and the presence of AGB stars in the selection creates an imaginary jump at $I = 26\,.\!\!^{\rm m}5$.
	
	The HST archive contains infrared images in the F110W and F160W filters obtained for the field at a distance of $30\arcmin$ from NGC\,1433. We carried out photometry of these images and found that an increase in the number of stars is observed at F160W = $25\,\!^{\rm m}$ (Fig.~\ref{fig10}). There are few such stars, but the jump in the luminosity function is clearly visible. If these are red giants, then the distance to them corresponds to the distance to NGC\,1433, since the difference between TRGB jumps in the F160W and F814W ($I$) filters is approximately $2\,\!^{\rm m}$. These stars can belong to the distant periphery of NGC\,1433, or be intergalactic stars. The angular distance between NGC\,1433 and this field corresponds to 124~kpc. In the massive elliptical galaxy M\,87, the stellar periphery is traced up to a distance of 190~kpc  \citep{Tikhonov_etal_2019}, and in the lenticular galaxy NGC\,5129, the luminosity of which is similar to the luminosity of NGC\,1433, the stellar halo is traced up to 140~kpc \citep{Rejkuba_etal_2014}. Thus, there is reason to believe that the halo of NGC\,1433 extends to 124~kpc.
	\normalem
\begin{acknowledgements}
	This study is based on observations with the NASA/ESA Hubble Space Telescope, obtained at the Space Telescope Science Institute, which is operated by AURA, Inc. under contract N\,NAS5-26555. These observations are associated with proposals 10438, 10354, 12999, 13364, 15654.
	In this work, we used NED and HyperLeda databases.
	
	The reported study was founded by RFBR and NSFB according to the research project  N\,19-52-18007/19.

\end{acknowledgements}


\begin{thebibliography}{99}
	
\bibitem[Bertelli~et~al.(1994)]{Bertelli_etal_1994}  G.~Bertelli, A.~Bressan, C.~Chiosi et al., Astron. Astrophys. {\bf106}, 275 (1994).

\bibitem[Carrasco~et~al.(2001)]{Carrasco_etal_2001} E.R.~Carrasco, C.~Mendes de Oliveira, L.~Infante, M.~Bolte, Astron. J., {\bf121}, 148 (2001).

\bibitem[Cattapan~et~al.(2019)]{Cattapan_etal_2019} A.~Cattapan, M.~Spavone, E.~Iodice et al., Astrophys. J., {\bf874}, 130 (2019).

\bibitem[DeGraaff et al.(2007)]{DeGraaff_etal_2007} R.B.~DeGraaff, J.P.~Blakeslee, G.R.~Meurer and M.E.~Putman, Astrophys. J., {\bf671}, 1624 (2007).

\bibitem[Dolphin(2016)]{Dolphin_2016} Dolphin A., DOLPHOT: Stellar photometry, Astrophysics Source Code Library ascl:1608.013 (2016).

\bibitem[Ferguson and Sandage(1990)]{Ferguson_Sandage_1990} H.C.~Ferguson and A.~Sandage,  Astron. J., {\bf100}, 1 (1990).

\bibitem[Firth~et~al.(2006)]{Firth_etal_2006} P.~Firth, E.~Evstigneeva, J.B.~Jones et al., Monthly Notic. of the Roy. Astron. Soc., {\bf372}, 1856 (2006).

\bibitem[Giraud(1985)]{Giraud_1985} E.~Girand, Astron. Astrophys., {\bf153}, 125 (1985).

\bibitem[Huchra and Geller(1982)]{Huchra_Geller_1982} J.P.~Huchra and M.J.~Geller, Astrophys. J., {\bf257}, 423 (1982).

\bibitem[Kilborn et al.(2005)]{Kilborn_etal_2005} V.A.~Kilborn, B.S.~Koribalski, D.A.~Forbes et al., Monthly Notic. of the Roy. Astron. Soc., {\bf356}, 77 (2005).

\bibitem[Lee~et~al.(1993)]{Lee_etal_1993} M.G.~Lee, W.L.~Freedman  and B.F.~Madore, Astrophys. J., {\bf417}, 553 (1993).

\bibitem[Madore and Freedman(1995)]{Madore_Freedman_1995} B.~Madore and W.~Freedman, Astron.J., {\bf109}, 1645 (1995).

\bibitem[Maia~et~al.(1989)]{Maia_etal_1989} M.A.G.~Maia, L.N.~da Costa and D.W.~Latham, Astrophys. J. Supp., {\bf69}, 809 (1989).

\bibitem[Rejkuba~et~al.(2014)]{Rejkuba_etal_2014} M.~Rejkuba, W.E.~Harris, L.~Greggio et al., Astrophys. J. Lett. {\bf791}, L2 (2014).

\bibitem[Ryan-Weber~et~al.(2004)]{Ryan-Weber_etal_2004} E.V.~Ryan-Weber, G.R.~Meurer, K.C.~Freeman et al., Astron. J., {\bf127}, 1431 (2004).

\bibitem[Sabbi~et~al.(2018)]{Sabbi_etal_2018} E.~Sabbi, D.~Calzetti, L.~Ubeda, A.~Adamo et al., Astrophys. J. Supp., {\bf235}, 23 (2018).

\bibitem[Sandage(1975)]{Sandage_1975} A.~Sandage, Astrophys. J. {\bf202}, 563 (1975).

\bibitem[Schlafly and Finkbeiner(2011)]{Schlafly_Finkbeiner_2011} E.F.~Schlafly and D.P.~Finkbeiner, Astrophys. J. {\bf737}, 103 (2011).

\bibitem[Shahbazyan(1957)]{SH_1957} R.K.~Shahbazian, Astronomicheskij Tsirkulyar, {\bf177}, 11 (1957). 

\bibitem[Sorce~et~al.(2014)]{Sorce_etal_2014} J.G.~Sorce, R.B.~Tully,H.M.~Courtois et al., Monthly Notic. of the Roy. Astron. Soc., {\bf444}, 527 (2014).

\bibitem[Springob~et~al.(2014)]{Springob_etal_2014} C.M.~Springob, J.~Mould,  C.~Magoulas et al., Monthly Notic. of the Roy. Astron. Soc., {\bf445}, 2677 (2014).

\bibitem[Stetson(1987)]{Stetson_1987} P.B.~Stetson, Publ. Astron. Soc. Pacific, {\bf99}, 191 (1987).

\bibitem[Stetson(1994)]{Stetson_1994} P.B.~Stetson, Publ. Astron. Soc. Pacific, {\bf106}, 250 (1994).

\bibitem[Tikhonov(2018)]{Tikhonov_2018} N.A.~Tikhonov, Astrophys. Bull., {\bf73}, 23 (2018).

\bibitem[Tikhonov~et~al.(2009)]{Tikhonov_etal_2009} N.A.~Tikhonov, O.A.~Galazutdinova and E.N.~Tikhonov, Astron. Lett. {\bf35}, 559 (2009).

\bibitem[Tikhonov and Galazutdinova(2009)]{Tikhonov_Galazutdinova_2009} N.A.~Tikhonov and O.A.~Galazutdinova, Astron. Lett., {\bf35}, 748 (2009).

\bibitem[Tikhonov~et~al.(2014)]{Tikhonov_etal_2014} N.A.~Tikhonov, O.A.~Galazutdinova and V.S.~Lebedev, Astron. Lett., {\bf40}, 1 (2014).

\bibitem[Tikhonov and Galazutdinova(2018)]{Tikhonov_Galazutdinova_2018} N.A.~Tikhonov and O.A.~Galazutdinova, Astrophys. Bull., {\bf73}, 279 (2018).

\bibitem[Tikhonov~et~al.(2019)]{Tikhonov_etal_2019} N.A.~Tikhonov, O.A.~Galazutdinova and G.M. Karataeva, Astrophys. Bull., {\bf74}, 257 (2019).

\bibitem[Tonry~et~al.(2001)]{Tonry_etal_2001} J.L.~Tonry, A.~Dressler, J.P.Blakeslee et al., Astrophys. J., {\bf546}, 681 (2001).

\bibitem[Tully and Fisher(1988)]{Tully_Fisher_1988} R.B.~Tully and J.R.~Fisher, Catalog of Nearby Galaxies, (1988).

\bibitem[Tully~et~al.(2009)]{Tully_etal_2009} R.B.~Tully, L.~Rizzi, E.J.~Shaya et al., Astron. J., {\bf138}, 323 (2009).

\bibitem[Tully~et~al.(2013)]{Tully_etal_2013} R.B.~Tully, H.M.~Courtois, A.E.~Dolphin et al., Astron. J., {\bf146}, 86 (2013).

\bibitem[Shobbrook(1966)]{Shobbrook_1966} R.R.~Shobbrook, Monthly Notic. of the Roy. Astron. Soc.{\bf131}, 365 (1966).

\bibitem[Vaucouleurs(1975)]{Vaucouleurs_1975}  G.~de Vaucouleurs, Stars and Stellar Systems, {\bf9}, 557 (1975).

\bibitem[Werk~et~al.(2008)]{Werk_etal_2008} J.K.~Werk, M.E.~Putman, G.R.~Meurer et al., Astrophys. J., {\bf678}, 888 (2008).

\bibitem[Werk~et~al.(2010)]{Werk_etal_2010} J.K.~Werk, M.E.~Putman, G.R.~Meurer et al., Astron. J., {\bf139}, 279 (2010).

\bibitem[Willick~et~al.(1997)]{Willick_etal_1997} J.A.~Willick, S.~Courteau, S.M.~Faber et al., Astrophys. J. Suppl., {\bf 109}, 333 (1997).

\end{thebibliography}
\end{document}